\documentclass[iop]{emulateapj}
\slugcomment{Accepted by ApJ, December 19, 2013} 
\usepackage{amsmath,hyperref,breakurl}

\shorttitle{Edge-On Orion Protostar}
\shortauthors{Fischer et al.}

\begin{document}

\title{HOPS 136: An Edge-On Orion Protostar Near the End of Envelope Infall}
\author{William J. Fischer\altaffilmark{1}, S. Thomas Megeath\altaffilmark{1}, John J. Tobin\altaffilmark{2,3}, Lee Hartmann\altaffilmark{4}, Amelia M. Stutz\altaffilmark{5},\\ Marina Kounkel\altaffilmark{4}, Charles A. Poteet\altaffilmark{6}, Babar Ali\altaffilmark{7}, Mayra Osorio\altaffilmark{8}, P. Manoj\altaffilmark{9}, Ian Remming\altaffilmark{10},\\ Thomas Stanke\altaffilmark{11}, and Dan M. Watson\altaffilmark{12}}
\altaffiltext{1}{Department of Physics and Astronomy, University of Toledo, Toledo, OH, USA; wfische@utnet.utoledo.edu}
\altaffiltext{2}{National Radio Astronomy Observatory, Charlottesville, VA, USA}
\altaffiltext{3}{Hubble Fellow}
\altaffiltext{4}{Department of Astronomy, University of Michigan, Ann Arbor, MI, USA}
\altaffiltext{5}{Max-Planck-Institut f\"ur Astronomie, Heidelberg, Germany}
\altaffiltext{6}{New York Center for Astrobiology, Rensselaer Polytechnic Institute, Troy, NY, USA}
\altaffiltext{7}{NHSC/IPAC/Caltech, Pasadena, CA, USA}
\altaffiltext{8}{Instituto de Astrof\'{i}sica de Andaluc\'{i}a, CSIC, Granada, Spain}
\altaffiltext{9}{Department of Astronomy and Astrophysics, Tata Institute of Fundamental Research, Mumbai, India}
\altaffiltext{10}{Department of Astronomy and Astrophysics, University of Chicago, Chicago, IL, USA}
\altaffiltext{11}{ESO, Garching bei M\"unchen, Germany}
\altaffiltext{12}{Department of Physics and Astronomy, University of Rochester, Rochester, NY, USA}

 
\begin{abstract}
Edge-on protostars are valuable for understanding the disk and envelope properties of embedded young stellar objects, since the disk, envelope, and envelope cavities are all distinctly visible in resolved images and well constrained in modeling.  Comparing 2MASS, {\em WISE}, {\em Spitzer}, {\em Herschel}, APEX, and IRAM photometry and limits from 1.25 to 1200~\micron, {\em Spitzer} spectroscopy from 5 to 40 \micron, and high-resolution {\em Hubble} imaging at 1.60 and 2.05~\micron\ to radiative transfer modeling, we determine envelope and disk properties for the Class~I protostar HOPS~136, an edge-on source in Orion's Lynds 1641 region.  The source has a bolometric luminosity of 0.8 $L_\sun$, a bolometric temperature of 170~K, and a ratio of submillimeter to bolometric luminosity of 0.8\%.  Via modeling, we find a total luminosity of 4.7~$L_\sun$ (larger than the observed luminosity due to extinction by the disk), an envelope mass of 0.06~$M_\sun$, and a disk radius and mass of 450 AU and 0.002~$M_\sun$. The stellar mass is highly uncertain but is estimated to fall between 0.4 and 0.5 $M_\sun$.  To reproduce the flux and wavelength of the near-infrared scattered-light peak in the spectral energy distribution, we require $5.4\times 10^{-5}~M_\sun$ of gas and dust in each cavity.  The disk has a large radius and a mass typical of more evolved T Tauri stars in spite of the significant remaining envelope.  HOPS 136 appears to be a key link between the protostellar and optically revealed stages of star formation.
\end{abstract}


\section{INTRODUCTION}\label{s.intro}

From the early example of HH 30 \citep{bur96}, the {\em Hubble Space Telescope} ({\em HST}) has enabled dramatic advances in the study of edge-on circumstellar disks associated with young stellar objects (YSOs).  In this favorable geometry, the distribution of light scattered by the disk allows good estimates of several disk properties that are not easily discernible from the modeling of spectral energy distributions (SEDs) alone \citep{wat07}.  Edge-on systems are especially valuable at the protostellar stage when an envelope is present, since the disk, envelope, and bipolar envelope cavities can appear as distinct features, making it easier to disentangle their contributions to the scattered-light emission.  Some investigators, e.g., \citet{pad99}, \citet{gra10}, and \citet{tob08,tob10a}, have modeled images and multiwavelength photometry of edge-on protostellar systems in the relatively nearby Taurus-Auriga region (at a distance of 140 pc; \citealt{ber06}) to determine their disk, envelope, and cavity properties.

Here we bring the superb angular resolution of {\em HST} to an edge-on source in the more distant Orion star-forming region, taken to be at 420 pc based on high-precision parallax studies of non-thermal sources in the Orion Nebula region \citep{san07,men07,kim08}.  The source was discovered in a {\em Spitzer Space Telescope} survey of the Orion A and B molecular clouds \citep{meg12}.  At $\alpha=5^h38^m46^s.54$, $\delta=-7^\circ05'37''.4$ (J2000), it is an isolated object in the Lynds 1641 (L 1641) region of Orion A \citep{all08}.  A search of the literature reveals no previous imaging studies of the object, although it is source 1224 in the study of \citet{fan13}, who analyzed its 2MASS and {\em Spitzer} magnitudes in a study of over 1000 YSOs in L 1641.  We refer to it here as HOPS 136, its number in the target catalog for HOPS, the {\em Herschel} Orion Protostar Survey \citep{fis13,man13,stu13}.  HOPS was a 200-hour open-time key program with the {\em Herschel Space Observatory}\footnote{{\em Herschel} is an ESA space observatory with science instruments provided by European-led Principal Investigator consortia and with important participation from NASA.} \citep{pil10} to obtain imaging, photometry, and spectroscopy of Orion protostars between 55 and 210~\micron, where their SEDs are expected to peak, with PACS, the Photodetector Array Camera and Spectrometer \citep{pog10}.

In addition to the {\em Herschel} data, HOPS features imaging and spectroscopy from near-infrared to millimeter wavelengths.  Imaging of HOPS 136 with the Near Infrared Camera and Multi-Object Spectrometer (NICMOS) on board {\em HST} revealed bipolar nebulosity divided by a dark lane, confirming its edge-on orientation.  Here we model the 1.2 \micron\ to 1200 \micron\ SED and near-IR images of the source with the radiative transfer code of \citet{whi03b}.  Combined image and SED modeling breaks degeneracies common in the modeling of SEDs alone, in particular that a system with a tenuous envelope seen through an edge-on disk can have a similar SED to a system with a dense envelope seen closer to pole-on \citep{whi03a} or to a transition disk with an inner hole \citep{rib13}.  Based on our modeling, we find a low-mass protostellar envelope (0.06 $M_\sun$), a large, low-mass disk (radius 450 AU, mass 0.002 $M_\sun$), and a small amount of cavity material ($5.4\times 10^{-5}~M_\sun$ per cavity) for the source.

The paper is organized as follows:\  Section 2 discusses the target selection and the acquisition of multiwavelength imaging, photometry, and spectra, Section 3 presents an initial characterization of the source, Section 4 describes the model and inferred disk, envelope, and cavity properties, Section 5 contains discussion, and Section 6, our conclusions.

\section{TARGET SELECTION AND OBSERVATIONS}

\begin{figure}
\includegraphics[bb = 65 307 495 730, width=\hsize]{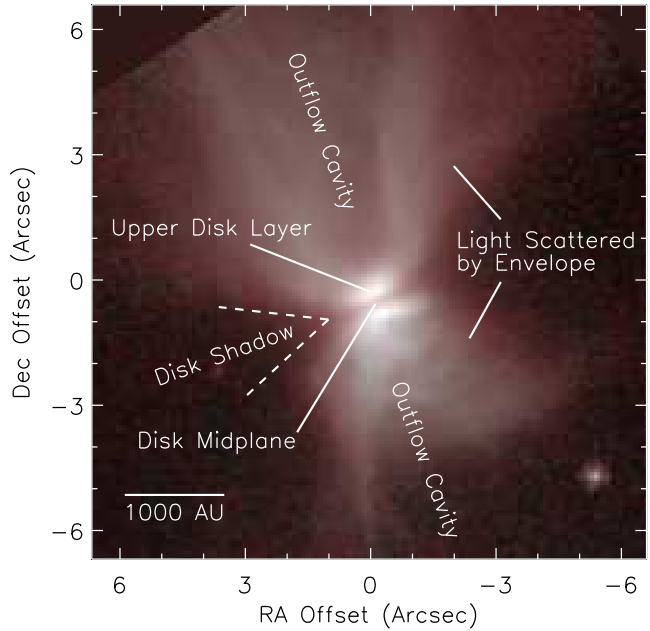}
\figcaption{Annotated two-color 1.60 \micron\ (cyan) and 2.05 \micron\ (red) NICMOS image of HOPS 136.  Offsets are from the {\em Spitzer}-determined J2000 position of the object ($\alpha=5^h38^m46^s.54$, $\delta=-7^\circ05'37''.4$).\label{f.nicmos}}
\end{figure}

In 2008 August and September, we acquired images of 72 objects in the HOPS catalog at 1.60 and 2.05 \micron\ with NICMOS.  Extended nebulosity and previously unknown companions are frequently detected (M.\ Kounkel et al., in preparation).  Five of the 72 sources have bipolar nebulosity divided by a dark lane, indicating an edge-on geometry.  From among these, we chose HOPS 136 for detailed modeling due to the full availability of photometry from 1.2~\micron\ to 1200~\micron\ and its almost exactly edge-on orientation (Section 3).  The combined 1.60 \micron\ and 2.05~\micron\ images of HOPS 136 appear in Figure~\ref{f.nicmos}.  The full set of images from the NICMOS survey and its extension with Wide Field Camera 3 (WFC3) will be presented by J.\ Booker et al.\ in a future publication.  Below, we provide details of the {\em HST} imaging, multiwavelength photometry, and {\em Spitzer} spectroscopy.  Table~\ref{t.photometry} contains the photometry for HOPS 136, and Figure~\ref{f.sed} shows its SED.

\subsection{HST Imaging\label{s.nicmos}}

We used NICMOS to map a field containing HOPS 136 on 2008 August 22.  We used the NIC2 camera, which has $256\times256$ pixels at 0.075\arcsec\ on a side for a 19.2\arcsec\ field of view, about 8000~AU at the distance of Orion.  The spatial resolution is about 80~AU.  The source was imaged with the F160W (1.60 \micron) and the F205W (2.05~\micron) filters.  In addition to the pipeline processing, we subtracted a median-combined image of a blank field to remove a spatially variable glow from the detector and telescope as well as variable offsets among the four detector quadrants.

\subsection{2MASS, WISE, and Spitzer Photometry}

We obtained photometry at $J$, $H$, and $K_s$ from the Two Micron All Sky Survey (2MASS; \citealt{skr06}), which observed HOPS 136 on 1998 March 30.  The source is flagged as extended, so we report fluxes within a circular aperture of radius 5\arcsec\ as given in the extended source catalog.  We also obtained photometry at 3.4, 4.6, 12, and 22~\micron\ from the {\em Wide-field Infrared Survey Explorer} ({\em WISE}; \citealt{wri10}), which observed the source on 2010 March 9--10.\footnote{Archival 2MASS and {\em WISE} photometry are available at \url{http://irsa.ipac.caltech.edu/}.}

Photometry at 3.6, 4.5, 5.8, 8.0, and 24 \micron\ for HOPS 136 was obtained as part of a joint survey of the Orion A and B molecular clouds by the Infrared Array Camera (IRAC; \citealt{faz04}) and Multiband Imaging Photometer (MIPS; \citealt{rie04}) on board {\em Spitzer}.  A detailed accounting of the {\em Spitzer} observations, data reduction, and source extraction can be found in \citet{kry12} and \citet{meg12}; here we summarize the most important details.

The IRAC observations of L 1641 were taken as part of Guaranteed Time Observation program 43 and were obtained in two epochs, one on 2004 February 17--19 and the other on 2004 October 8 and 27.  IRAC photometry was obtained at 3.6, 4.5, 5.8, and 8.0~\micron\ using an aperture of radius 2.4\arcsec\ with a sky annulus extending from 2.4\arcsec\ to 7.2\arcsec\ and corrected to infinity by dividing by aperture corrections:\ 0.824, 0.810, 0.725, and 0.631 in order of increasing wavelength.  Photometric zero points and zero-magnitude fluxes for the IRAC bands can be found in \citet{meg12} and \citet{rea05}, respectively.  The error estimate is dominated by calibration uncertainties, which we estimate to be 5\% in all channels.

The MIPS observations of L 1641 were taken as part of Guaranteed Time Observation program 47 on 2005 April 2--3.  MIPS photometry was obtained at 24~\micron\ by fitting a point-spread function \citep{kry12}.  The error estimate is dominated by a 5\% calibration uncertainty.

\begin{deluxetable}{ccccc}
\tablecaption{Photometry for HOPS 136\label{t.photometry}}
\tablewidth{\hsize}
\tablehead{\colhead{$\lambda$} & \colhead{$F_\nu$} & \colhead{$\sigma F_\nu$} & \colhead{Instrument} & \colhead{Date} \\ \colhead{(\micron)} & \colhead{(mJy)} & \colhead{(mJy)} & \colhead{} & \colhead{}}
\startdata
1.2  & 0.872 & 0.152 & 2MASS & 1998 Mar 30 \\
1.7  & 4.37 & 0.221 & 2MASS & 1998 Mar 30 \\
2.2  & 9.23 & 0.281 & 2MASS & 1998 Mar 30 \\
3.4  & 8.90 & 0.246 & $WISE$ & 2010 Mar 9--10 \\
3.6  & 8.83 & 0.443 & IRAC & 2004 Feb 17--Oct 27 \\
4.5  & 8.77 & 0.440 & IRAC & 2004 Feb 17--Oct 27 \\
4.6  & 10.9 & 0.270 & $WISE$ & 2010 Mar 9--10 \\
5.8  & 7.16 & 0.361 & IRAC & 2004 Feb 17--Oct 27 \\
8.0  & 4.51 & 0.229 & IRAC & 2004 Feb 17--Oct 27 \\
12  & 4.20 & 0.259 & $WISE$ & 2010 Mar 9--10 \\
22  & 59.9 & 3.20 & $WISE$ & 2010 Mar 9--10 \\
24 & 64.6 & 3.33 & MIPS & 2005 Apr 2 \\
70  & 1670 & 83.8 & PACS & 2010 Sep 28 \\
100 & 2030 & 108 & PACS & 2011 Apr 11 \\
160 & 2110 & 115 & PACS & 2010 Sep 28 \\
350 & 423 & 169 & SABOCA & 2010 Aug 18 \\
870 & 34.0 & 6.81 & LABOCA & 2010 Oct 22 \\
1200 & $<36$ & \nodata & MAMBO & 1999 Feb
\enddata
\end{deluxetable}

\begin{figure}
\includegraphics[bb = 72 290 560 685, width=\hsize]{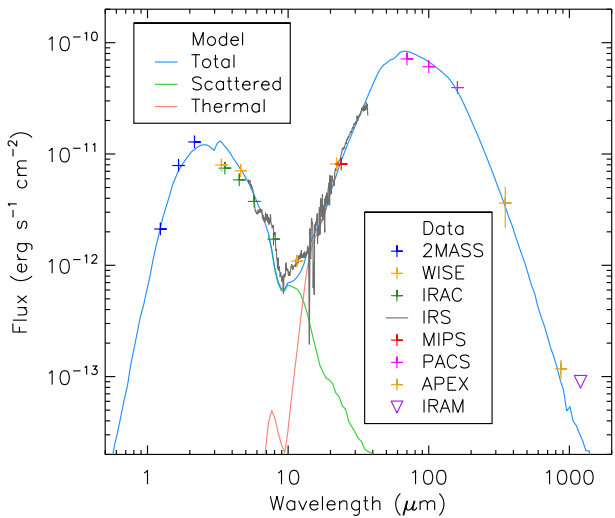}
\figcaption{Photometry and IRS spectrum of HOPS 136 with model fit.  The triangle indicates an upper limit.  The total model, discussed in Section 4, is shown as well as its decomposition into scattered-light and thermal components, which dominate the short and long wavelength peaks, respectively.\label{f.sed}}
\end{figure}

\vskip 0.5 in

\subsection{Herschel/PACS Photometry}

With {\em Herschel} we observed a $5'$ square field centered on HOPS 136 on 2010 September 28 (observing day 502; observation IDs 1342205242 and 1342205243) in the 70~$\mu$m and 160 $\mu$m bands available with PACS, which have angular resolutions of 5.2\arcsec\ and 12\arcsec, respectively.  We observed our target field with homogeneous coverage using two orthogonal scanning directions and a scan speed of 20\arcsec\ s$^{-1}$.  Each scan was repeated six times for a total observation time of 481~s per scan direction.  For photometry, the {\em Herschel} data were processed with the high-pass filtering method described by \citet{fis10} and discussed in detail by \citet{pop12}, using version 8.0, build 248 of HIPE, the {\em Herschel} Interactive Processing Environment \citep{ott10}.

HOPS 136 is unresolved at both PACS wavelengths.  We obtained aperture photometry using a 9.6\arcsec\ radius at 70 \micron\ and a 12.8\arcsec\ radius at 160 \micron\ with subtraction of the median signal in a background annulus extending from the aperture limit to twice that value in both channels.  The results were divided by measurements of the encircled energy fractions in these apertures provided by the {\tt photApertureCorrectionPointSource} task in HIPE \citep{lut12}, adjusted for the fact that our close-in sky subtraction removes 3\%--4\% of the flux in each point-spread function.  These aperture corrections are 0.733 at 70 \micron\ and 0.660 at 160 \micron.  The error estimate is dominated by calibration uncertainties \citep{bal13}, which we estimate to be 5\% at both wavelengths.

To supplement the far-IR SED coverage, we also include a 100 \micron\ measurement from the {\em Herschel} Gould Belt Survey \citep{and10}.  From the {\em Herschel} Science Archive\footnote{The {\em Herschel} Science Archive can be accessed online at \url{http://herschel.esac.esa.int/Science\_Archive.shtml}.}, we obtained two orthogonal scans of a region containing HOPS 136 (observation IDs 1342218553 and 1342218554), and we reduced them with the same high-pass filtering method we used for the 70 and 160~\micron\ data.  We obtained aperture photometry for HOPS 136 using a 9.6\arcsec\ radius with subtraction of the median signal in a background annulus extending from 9.6\arcsec\ to 19.2\arcsec, the same as at 70 \micron, and we divided by an aperture correction of 0.694, again accounting for removal of some of the point-spread function by sky subtraction.  We assume the same 5\% error floor as for the other PACS bands.

\subsection{(Sub)millimeter Photometry}

In 2010, we observed HOPS 136 with APEX, the Atacama Pathfinder Experiment.  We acquired a 350~\micron\ image on August 18 with SABOCA, the Submillimetre APEX Bolometer Camera \citep{sir10}, and we acquired an 870 \micron\ image on October 22 with LABOCA, the Large APEX Bolometer Camera \citep{sir09}.  Details of the reduction and calibration can be found in \citet{fis12} and \citet{stu13}, and the data will be discussed in more detail by T.\ Stanke et al.\ (in preparation).  As in \citet{fis12}, for photometry we report the flux densities without background subtraction in circular apertures of diameter equal to the FWHM of the instrument beam (7.3\arcsec\ at 350 \micron\ and 19\arcsec\ at 870 \micron).  The flux calibration is accurate to within 40\% for SABOCA and 20\% for LABOCA.

A dust continuum map of the HOPS 136 region at 1.2~mm was obtained in 1999 February with the 37 channel MAMBO bolometer array at the IRAM 30 m telescope.  The data were reduced with standard procedures for bolometers as described in \citet{sta07}.  At this wavelength, HOPS 136 was undetected, and we present an upper limit.

\subsection{Spitzer/IRS Spectroscopy}

We observed HOPS 136 (target 8469384-7093; AOR key 20856320) on 2007 April 16 with the {\em Spitzer} Infrared Spectrograph (IRS; \citealt{hou04}). The observations were made with the two low-spectral-resolution IRS modules (short-low from 5.2 to 14.5 \micron\ and long-low from 14.0 to 38.0 \micron; $\lambda/\Delta\lambda=60-120$) in staring mode. The spectrum was generated from the {\em Spitzer} Science Center (SSC) S18.7 pipeline basic calibrated data with the IRS instrument team's SMART software package \citep{hig04}. To prepare the data for extraction, we first replaced the permanently bad and ``rogue'' pixels' flux values with those interpolated from neighboring functional pixels.  For the short-low module and the first order of the long-low module, sky emission was removed by subtracting the extracted profiles of off-nod or off-order data. The sky emission in the second and third orders of the long-low module was removed by subtracting a degree-zero polynomial that was fit to the emission profiles.

The spectra were extracted with the advanced optimal extraction method (AdOpt; \citealt{leb10}). We then used AdOpt again to extract the spectra of three spectral calibrators: Markarian 231, $\alpha$ Lac, and $\xi$ Dra. Template spectra of these three calibrators were divided by their extracted spectra at the two nod positions to create relative spectral response functions (RSRFs). We then multiplied the extracted orders of HOPS 136 at both nod positions by the RSRFs. The resulting nod-position spectra were averaged to obtain the final spectrum, and the spectral uncertainties are estimated to be half the difference between the two independent spectra from each nod position.

\section{SOURCE CHARACTERIZATION}\label{s.source}

Here we present some initial characterization of HOPS 136 before applying radiative transfer modeling.  Figure~\ref{f.nicmos} shows how the elements of a protostar except for the central source are all visible in scattered-light imaging of an edge-on system, with the dark disk midplane, bright upper disk layers, disk shadow, envelope, and cavities all denoted.

The near-IR morphology confirms an edge-on line of sight, and the brightness ratio of the two sides of the nebula is sensitive to the exact inclination \citep{wat07}.  To avoid complications from the complex envelope morphology, we measured the brightness in two boxes that allow a comparison of the upper and lower halves of the disk.  The boxes have widths of 31 pixels (980 AU) and extend above or below the central dark lane with heights of 8 pixels (250 AU).  The top box is 19\% brighter than the bottom box at 1.60 \micron\ and 4\% brighter than the bottom box at 2.05 \micron.  In the model adopted below, a system even 1$^\circ$ from edge-on has 24\% more flux in the upper box at 1.60 \micron\ and 22\% more flux at 2.05 \micron.  While inclination is not the only factor that influences the brightness ratio, this finding suggests that the inclination of HOPS 136 is greater than $89^\circ$.  In the modeling below, we fix the inclination at $90^\circ$.

We treat the SED as static due to the lack of significant variability in mid-IR photometry and spectra acquired in 2004, 2005, 2007, and 2010. The bolometric luminosity and temperature derived from a protostellar SED characterize its evolutionary state \citep{mye93}.  For HOPS 136, the bolometric luminosity $L_{\rm bol}$, found by trapezoidal integration of the SED and assuming a distance of 420 pc to the Orion region, is 0.83 $L_\sun$.  The bolometric temperature $T_{\rm bol}$, the temperature of a blackbody with the same mean frequency as the SED, is 174~K.  This places HOPS 136 squarely in the observational category Class I, with $70~{\rm K}<T_{\rm bol}<650~{\rm K}$ \citep{che95}.  Class I is believed to correspond, generally, to the evolutionary category Stage I \citep{dun13}, where the envelope material is in the process of falling onto a circumstellar disk but the stellar mass exceeds the remaining envelope mass.  The ratio of submillimeter to bolometric luminosity, $L_{\rm smm}/L_{\rm bol}$, where the APEX data are used to calculate $L_{\rm smm}$, is 0.8\%. This is greater than the 0.5\% upper limit for Class I sources \citep{and93}, implying a less evolved envelope, but in Section 5.3 we conclude that this is due to the edge-on inclination of the source rather than an atypically dense envelope for Stage I.  (The 1.2 mm MAMBO limit was ignored in these calculations, but treating it otherwise affects the results at a level far below the quoted precision.)

Analyzing the slope of the SED over various wavelength intervals, where \begin{equation}\alpha_{\lambda_1-\lambda_2}=\log \left(\lambda_1 S_{\lambda_1}/\lambda_2 S_{\lambda_2} \right) / \log \left(\lambda_1/\lambda_2\right)\end{equation} and $S_{\lambda_i}$ is defined as the flux density at wavelength $\lambda_i$, we find $\alpha_{2.2-24}=-0.2$, typical of the flat-SED sources \citep{gre94}.  These can be modeled as young stellar objects with envelopes \citep{cal94}.  \citet{fan13} drew the same conclusion about HOPS 136 (their object 1224) from $\alpha_{3.6-24}=0.05$.  In the system developed by \citet{mcc10} for IRS spectra, $\alpha_{5-12}=-1.6$ and $\alpha_{12-20}=3.7$, which would classify HOPS 136 as a transition disk, a post-protostellar object in which the protoplanetary disk has an inner hole.  Due to the deep local minimum centered near 10 \micron\ but affecting fluxes across the {\em Spitzer} range, it is necessary to compare the near-infared to the far-infrared to find the rising SED typical of Class I sources.  We find $\alpha_{2.2-70}=0.5$ for HOPS 136, consistent with its Class I status.

The source is in an isolated environment, with minimal nebular background and only three 70 \micron\ point sources, all at least ten times fainter, detected within $5\times10^4$ projected AU.  Further, HOPS 136 does not appear to be driving a jet.  While the cavities visible in the NICMOS images suggest outflow at some level, there are no affiliated Herbig-Haro objects, there are no outflow lines in its {\em Spitzer} spectrum, and an outflow is only marginally detected in a CO $3\rightarrow2$ map of this region (J.\ Di Francesco, priv.\ comm.).  This source provides an opportunity to study the formation of an isolated protostar, allowing a comparison to standard models.

The NICMOS images do show a faint point source 6.6\arcsec\ (projected separation 2800 AU) to the southwest of the protostar (Fig.~\ref{f.nicmos}).  This source, undetected by 2MASS, has NICMOS magnitudes $19.441\pm0.026$ in F160W and $18.527\pm0.032$ in F205W.  While this is likely a background star, if it is a physical companion, its magnitudes are consistent with those of an early L dwarf behind $A_V\sim5$ mag of extinction, according to the BT-Settl models for an age of 1 Myr \citep{all12}.  The scattered-light nebulosity is less extended in this direction, raising the possibility that the faint source influences the morphology of the HOPS 136 envelope.

\section{RADIATIVE TRANSFER MODELING}\label{s.model}

To gain a more precise understanding of the HOPS 136 envelope, disk, and central star, we used the Monte Carlo radiative transfer code of \citet{whi03b} to fit the source SED and images.  The code features a central star and flared disk, which emit photons that can then be scattered or absorbed and re-emitted by dust in either the disk or an envelope.  The envelope density is defined by the rotating collapse solution of \citet{ulr76} plus a bipolar cavity that can contain dust and gas with a power-law density distribution.

We first attempted to fit the SED of HOPS 136 with the \citet{rob07} online tool, which finds the best match to an input SED in a pre-computed grid of over 20,000 Whitney et al.\ models observed at 10 viewing angles \citep{rob06}.  Of the ten best-fit SEDs, three have viewing angles $>80^\circ$, consistent with the edge-on morphology of the NICMOS images.  These SEDs match the 1.2--1200 \micron\ photometry except for a mild underprediction of the flux near the 10~\micron\ minimum in the IRS spectrum.  However, the near-IR peak in all of these models is due to direct stellar flux, not scattered light.  In these models, the view toward the protostar skims the upper edge of the disk, allowing the near-IR flux of the central star to escape to the observer with little attenuation.  Since the Robitaille grid does not include images, we generated our own with the parameters of the best-fit SEDs.  These images contain a bright point source at the center of the system, not the dark dust lane observed toward HOPS 136.  No model in the Robitaille grid fits both the SED and the image well.

\begin{figure}
\includegraphics[bb = 72 370 560 710, width=\hsize]{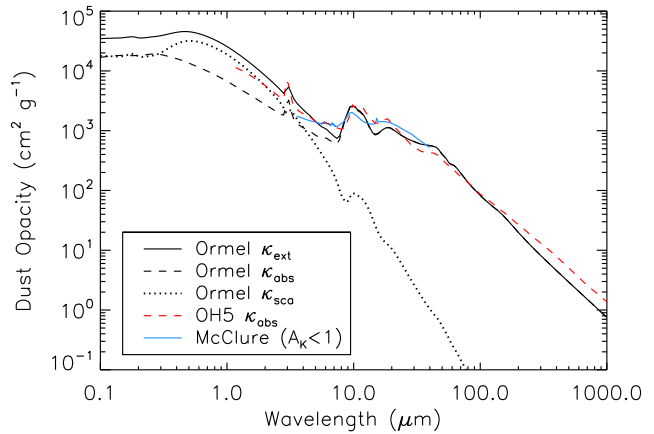}
\figcaption{Dust properties. Black curves show the adopted model from \citet{orm11}, with a dashed curve for absorption, a dotted curve for scattering, and a solid curve for their sum.  For comparison, the dashed red curve shows the IR ``OH5'' opacity from \citet{oss94}, and the solid blue curve shows the \citet{mcc09} mid-IR extinction for $A_K<1$. (\citeauthor{mcc09} tabulates $A_\lambda/A_K$; for presentation, we scale this to match the \citeauthor{orm11} opacity at 5 \micron.)\label{f.dust}}
\end{figure}

For subsequent fitting, we used the version of the Whitney et al.\ code with release date 2008 April 7.  For the disk, envelope, and cavity dust, we use a model from \citet{orm11} for a 2:1 mixture of ice-coated silicates and bare graphite grains, where the depth of the ice coating is 10\% of the particle radius.  The particles are subjected to time-dependent coagulation; we choose a time of 0.3 Myr.  The particle size distribution at this time ranges from $a=0.1$ to 3 \micron\ with the number at each size roughly proportional to $a^{-2.3}$.  The opacity and scattering laws for the ensemble are density-weighted averages of laws for particles of different sizes.  We assume a gas-to-dust ratio of 100.  The adopted model is shown in black in Fig.~\ref{f.dust}.  We favor this model because it contains both scattering and absorption properties across the full range of wavelengths required by the radiative transfer code, its mid-IR properties resemble those determined by \citet{mcc09} for star-forming regions (blue curve in Fig.~\ref{f.dust}; this is also what we use to add foreground reddening to the model) and those used by \citet{tob08} for modeling an edge-on protostar, and its mid-to-far-IR opacities resemble those of the frequently cited OH5 opacities (\citealt {oss94}; red dashed curve in Fig.~\ref{f.dust}).

The adjustable input parameters in the Whitney et al.\ code were initially set to those of the ``Class I'' model from \citet{whi03a}, which are typical values for low-mass protostars, and the inclination was set to $90^\circ$ due to the morphology of the NICMOS images.  This model yields a double-peaked SED, with a near-to-mid-infrared peak consisting of stellar and accretion-generated photons that have been scattered by the disk and envelope and a far-infrared peak consisting of photons from the same initial sources that have been reprocessed to long wavelengths by cold disk and envelope dust.  It also yields an image with two roughly parabolic nebulae separated by a dark lane.

The SED is most sensitive to the total luminosity $L$, which controls the overall flux in the SED, and the envelope density, parameterized here by the reference density $\rho_1$, the density at 1 AU in the limit of no rotation, which controls the wavelength and flux of the far-infrared peak \citep{ken93}.  These are not set explicitly in the Whitney et al.\ code but are functions of the infall and accretion rates and the stellar parameters.

The luminosity is the sum of the stellar luminosity $L_*$ and the accretion luminosity $L_{\rm acc}$, where $L_*\propto R_*^2~T_*^4$ and $L_{\rm acc}$ is a function of the stellar mass $M_*$, the stellar radius $R_*$, the inner radius of the dust disk $R_{\rm min, disk}$, the inner radius of the gas disk $R_{\rm trunc}$, and the rate at which matter accretes from the disk onto the star $\dot{M}_{\rm disk}$.  Since reprocessing by the disk and envelope erases the distinction between luminosity from the central source and luminosity due to accretion, we simplify the fitting by fixing the stellar luminosity at 1 $L_\sun$ and adjusting the total luminosity only via the disk-to-star accretion rate.

The envelope density is proportional to $\dot{M}_{\rm env}$, the rate at which matter falls from the envelope onto the disk.  For the adopted density profile, \citet{ken93} showed that the reference density $\rho_1$ can be written as
\begin{equation}
\begin{split}
\rho_1 = 7.5\times10^{-15}\left(\frac{\dot{M}_{\rm env}}{10^{-6}~M_\sun~{\rm yr}^{-1}}\right)\left(\frac{M_*}{0.5~M_\sun}\right)^{-1/2} \\ {\rm g~cm}^{-3}\label{e.rho1}.
\end{split}
\end{equation} As the density increases, the far-IR flux first increases and then shifts to longer wavelengths.  The first step in fitting was to coarsely adjust $L$ and $\rho_1$ to get an approximate match to the observed SED.

\begin{figure*}
\includegraphics[width=\hsize]{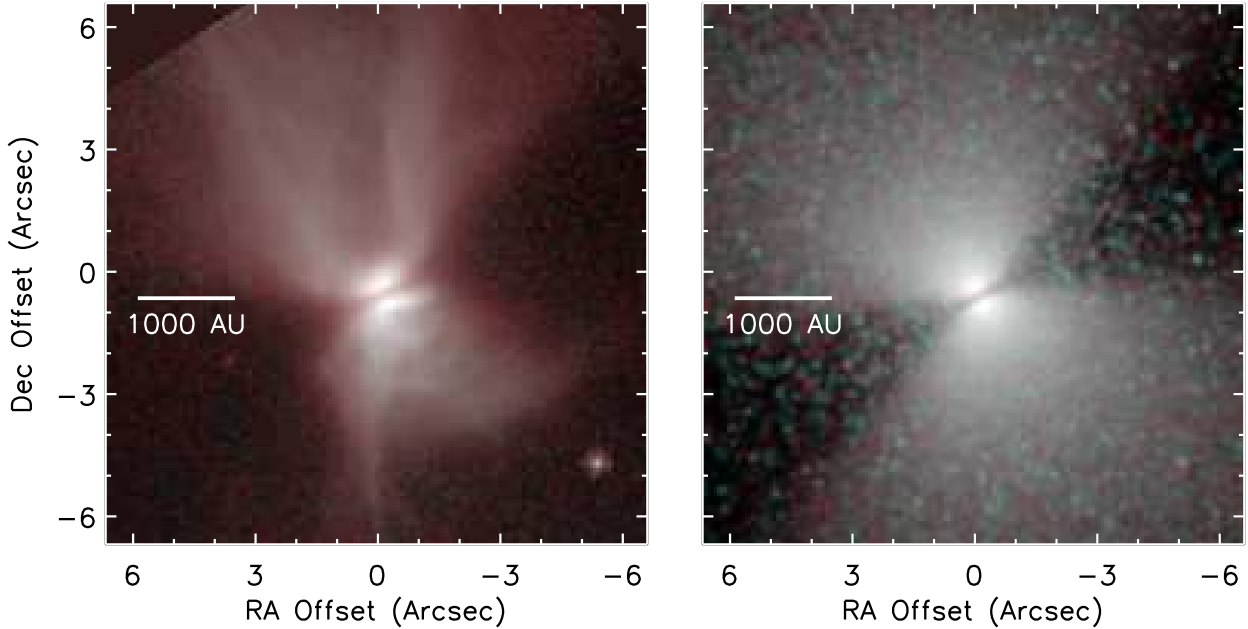}
\figcaption{{\em Left:} Two-color 1.60 \micron\ (cyan) and 2.05 \micron\ (red) NICMOS image of HOPS 136.  Offsets are from its {\em Spitzer}-determined J2000 position ($\alpha=5^h38^m46^s.54$, $\delta=-7^\circ05'47''.46$).  {\em Right:} Model image of HOPS 136 with the same color scheme.  The model output is convolved with a gaussian beam of FWHM $=0.2$\arcsec\ and then centered and rotated to match the orientation of the observed nebulosity.\label{f.model}}
\end{figure*}

To fit the images, we compared the height and width of the contours at 1, 10, and 35\% of the peak emission in the model to those in the observations.  These contours track the width and depth of the dark lane, the extent of the adjacent bright concentrations of scattered light, and the morphology of the envelope structure on a scale of $\sim1000$~AU, respectively.  Since our code is axisymmetric, we do not attempt to reproduce even fainter emission at larger scales, which is irregular in HOPS 136 and protostars in general \citep{tob10b}. 

After obtaining coarse fits to the SED and images by adjusting $L$ and $\rho_1$, we built grids of models to cover small ranges of parameters around the coarsely determined values.  We ranked the SED models with the $R$ statistic (\citealt{fis12}; E.\ Furlan et al., in preparation). This statistic measures the logarithmic deviation of the observed SED from the models in units of the fractional uncertainty, where \begin{equation}R=\sum_{i=1}^N\left[w_i\left| \ln \left(F_{\lambda_i{\rm, o}}/F_{\lambda_i{\rm, m}}\right)\right|\right]/N.\end{equation}  Here $N$ is the number of data points, $F_{\lambda_i{\rm, o}}$ is the observed flux at each wavelength $\lambda_i$, $F_{\lambda_i{\rm, m}}$ is the model flux at each wavelength $\lambda_i$, and $w_i$ is the inverse of the approximate fractional uncertainty in each data point, taken to be 5\% at wavelengths less than 350 \micron, 40\% at 350 \micron, and 20\% at 870 \micron .  Models that violate the 1200 \micron\ upper limit were discarded.  Each model can be shifted slightly in luminosity and modified with foreground reddening under the law of \citet{mcc09} to improve the fit (Section 4.1). The models with $R\lesssim4$ are qualitatively good fits, and we chose the one in this range that provides the best match to the image contours; it has $R=3.31$.

We compare the preferred model SED to the photometry and spectrum in Figure~\ref{f.sed}, and we compare the model images to the NICMOS images in Figure~\ref{f.model}.  To get adequate signal in the image and the millimeter-wavelength portion of the SED, $8\times10^7$ photons were run through the Monte Carlo code.  The code generates output for multiple apertures; the plotted SED shows the result from the 5\arcsec\ aperture in the 2MASS regime and the 20\arcsec\ aperture (which captures the entire simulation box) at longer wavelengths, with an interpolation scheme to bridge the change of aperture.  Counting all the flux in the simulation box is a good approximation to the observed non-2MASS fluxes, which are either measured in small apertures and corrected to a total flux via aperture corrections or determined by fitting a point-spread function.   

\begin{figure*}
\includegraphics[bb = 65 575 558 782, width=\hsize]{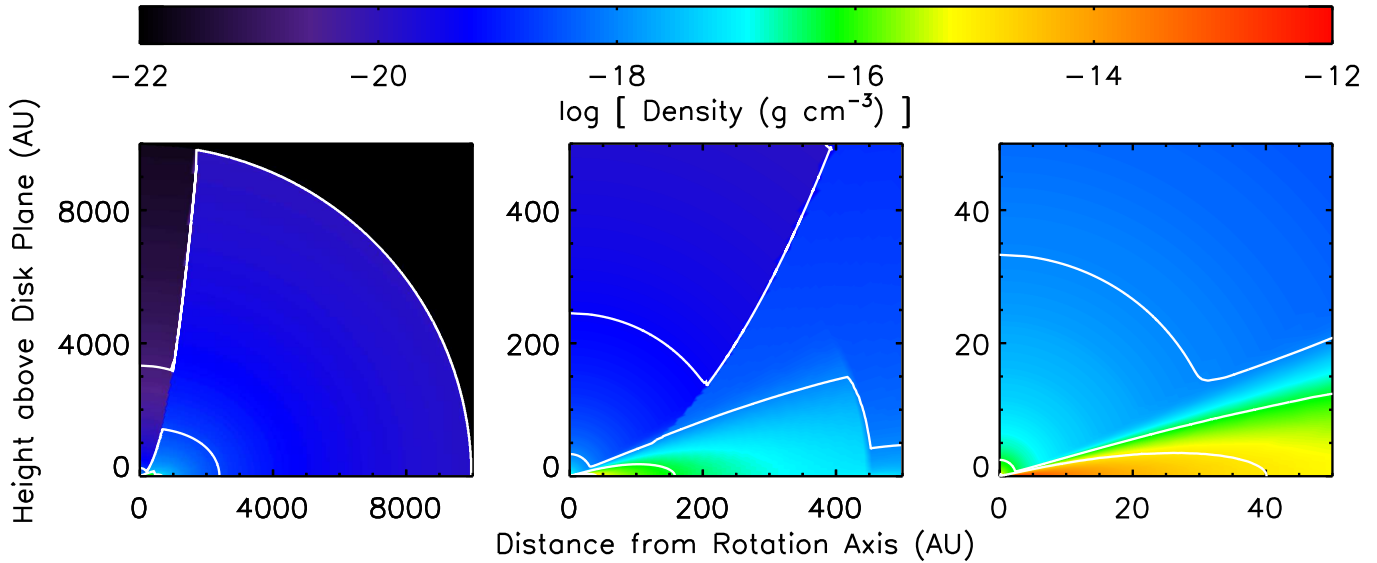}
\figcaption{Density distribution for the adopted model. From left to right, panels are scaled to emphasize the envelope, disk, and inner region.  Contours are at $10^{-21}$, $5\times10^{-20}$, $10^{-18}$, $5\times10^{-17}$, and $10^{-15}$ g cm$^{-3}$.\label{f.rho}}
\end{figure*}

\begin{figure*}
\includegraphics[bb = 65 575 558 782, width=\hsize]{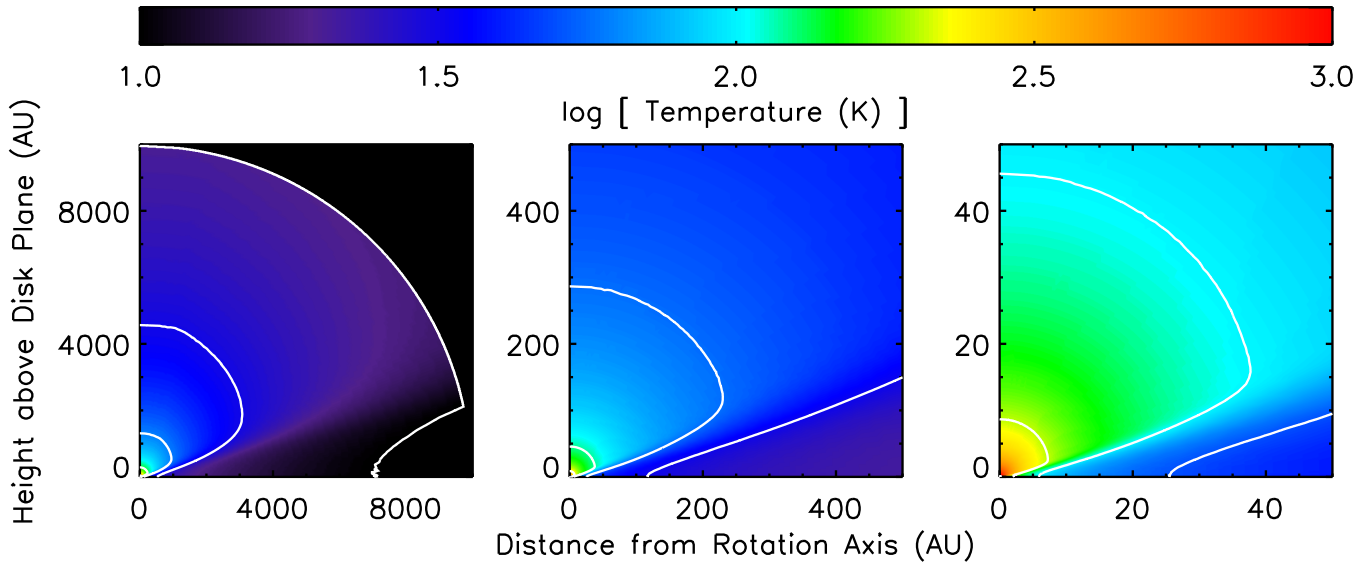}
\figcaption{Temperature distribution for the adopted model. From left to right, panels are scaled to emphasize the envelope, disk, and inner region.  Contours are at 10, 20, 30, 50, 100, and 200 K.\label{f.temp}}
\end{figure*}

We present the well constrained parameters of the adopted model and their uncertainties in Table~\ref{t.mainpars}.  For reproducibility, we include a comprehensive list of other input parameters in Table~\ref{t.otherpars}.  Both tables include italicized quantities of interest that are not directly specified but are instead derived from the input parameters, such as the total luminosity of the system.  Finally, the density and temperature distributions for the adopted model are presented in Figures~\ref{f.rho} and \ref{f.temp}.  In the following subsections, we discuss the well constrained parameters.  

\begin{deluxetable*}{llr}
\tablecaption{Well Constrained Properties of the Best-Fit Model\label{t.mainpars}}
\tablewidth{\hsize}
\tablehead{Property & Description & Value}
\startdata
$M_{\rm disk}$ ($M_\sun$) & Disk mass & $(2\pm0.5)\times10^{-3}$ \\
$h_{100}$ (AU) & {\em Disk scale height at 100 AU} & $12\pm2$ \\
$R_{\rm max, disk}$ (AU) & Outer disk radius & $450\pm25$ \\
$\rho_1$ (g cm$^{-3}$) & {\em Envelope density at 1 AU in the limit of no rotation} & $(4.5\pm0.5)\times10^{-15}$ \\
$\rho_{1000}$ (g cm$^{-3}$) & {\em Envelope density at 1000 AU in the limit of no rotation} & $(1.4\pm0.2)\times10^{-19}$ \\
$M_{\rm env}$ ($M_\sun$) & {\em Envelope mass inside 10,000 AU radius} & $0.060\pm0.005$ \\
$\theta_{\rm cav}$ ($^\circ$) & Cavity opening angle & $10\pm2$ \\
$\gamma_{\rm cav}$ & Cavity opening exponent & $2\pm0.5$ \\
$M_{\rm cav}$ ($M_\sun$) & {\em Mass in each cavity} & $(5.4\pm0.8)\times10^{-5}$ \\
$L$ ($L_\sun$) & {\em System luminosity} & $4.7\pm0.1$ \\
$i$ ($^\circ$) & Inclination angle & $90\pm1$ \\
$A_K$ (mag) & Foreground extinction at $K$ & $0.55\pm0.02$ 
\enddata
\tablecomments{Italicized properties are derived from the input parameters rather than specified directly.}
\end{deluxetable*}

\begin{deluxetable*}{llr}
\tablecaption{Other Input Parameters of the Best-Fit Model\label{t.otherpars}}
\tablewidth{\hsize}
\tablehead{Property & Description & Value}
\startdata
$R_*$ ($R_\sun$) & Stellar radius & 2.09 \\
$T_*$ (K) & Stellar temperature & 4000 \\
$M_*$ ($M_\sun$) & Stellar mass & 0.5 \\
$R_{\rm min, disk}$ ($R_*$) & {\em Inner dust disk radius = dust sublimation radius} & 14.8 \\
$h_0$ ($R_*$) & Disk scale height at $R_*$ (disk does not actually extend to $R_*$) & 0.02 \\
$\alpha$ & Disk radial density exponent & $-2.20$ \\
$\beta$ & Disk scale height exponent & 1.20 \\
$\dot{M}_{\rm disk}$ ($M_\sun~{\rm yr}^{-1}$) & Disk accretion rate & $6.7\times10^{-7}$ \\
$R_{\rm trunc}$ ($R_*$) & Inner gas disk radius & 3.0 \\
$f_{\rm spot}$ & Fractional area of accretion hot spots & 0.01 \\
$R_{\rm min, env}$ ($R_*$) & {\em Inner envelope radius = dust sublimation radius} & 14.8 \\
$R_C$ (AU) & Envelope centrifugal radius & 450 \\
$R_{\rm max, env}$ (AU) & Outer envelope radius & 10,000 \\
$\dot{M}_{\rm env}$ ($M_\sun~{\rm yr}^{-1}$) & Envelope infall rate & $6.0\times10^{-7}$ \\
$\eta_{\rm cav}$ & Cavity density exponent & $-1.5$ \\
$\rho_{0, \rm cav}$ (g cm$^{-3}$) & Density of cavity at $R_*$ (cavity material does not actually extend to $R_*$) & $2.0\times10^{-13}$ \\
$z_{0,\rm cav}$ (AU) & Offset of cavity base from midplane & 0 \\
$\rho_{\rm amb}$ (g cm$^{-3}$) & Ambient density & 0
\enddata
\tablecomments{These properties are tabulated for reproducibility and do not, by themselves, have a significant effect on the SED and image, although they may appear in the derivation of italicized properties in Table~\ref{t.mainpars} that are well constrained. Italicized properties are derived from the input parameters rather than specified directly.}
\end{deluxetable*}

\subsection{Luminosity and Foreground Extinction}

The overall system properties that are well constrained are the inclination, the luminosity, and the amount of foreground extinction.  As discussed in Section~\ref{s.source}, the inclination appears to be nearly edge-on, $90\pm1^\circ$.  

For an edge-on protostar, the intrinsic luminosity is generally larger than the observed luminosity due to extinction by the disk and by foreground dust.  In contrast to the observed luminosity of 0.8 $L_\sun$, the best-fit model has an intrinsic luminosity near 5 $L_\sun$ when all viewing angles are accounted for.  To optimize the luminosity and extinction, we assume that for small differences in luminosity, SEDs differ by only a multiplicative constant.  This is reasonable given the finding of \citet{ken93} that the wavelength of the peak flux from an optically thick envelope scales as the luminosity to only the $-1/12$ power.  We then write the observed SED $O_\lambda$ as \begin{eqnarray}O_\lambda&=&\ell M_\lambda 10^{-0.4A_\lambda}\nonumber\\
&=&\ell M_\lambda10^{-0.4 A_K k_\lambda},
\end{eqnarray}
where $M_\lambda$ is a model SED, $\ell$ is a small multiplicative constant that allows the luminosity to differ slightly from that of the best-fit model, $A_\lambda$ is the foreground extinction in magnitudes as a function of wavelength, $A_K$ is the extinction in the $K$ band, and the extinction law is $k_\lambda=A_\lambda/A_K$.  We use the extinction law of \citet{mcc09}, plotted in Figure~\ref{f.dust}.  Rearranging terms,
\begin{equation}2.5\log\left(M_\lambda/O_\lambda\right)=A_K k_\lambda-2.5\log \ell.\end{equation}This equation is linear in $k_\lambda$.  The plot of $2.5\log\left(M_\lambda/O_\lambda\right)$ versus $k_\lambda$ can then be fit with a line having slope $A_K$ and intercept $-2.5\log \ell$.  Following this procedure, we find a luminosity $4.7\pm0.1~L_\sun$ for the system and a foreground extinction of $A_K=0.55\pm0.02$ mag.

\vskip 0.5 in

\subsection{Disk Properties}

According to \citet{wat07}, the properties of an edge-on disk best constrained by modeling are the mass-opacity product of the disk and the effective scale height of dust in the outer disk.  The mass-opacity product is constrained rather than the mass alone, since the mass density and the opacity per unit mass appear as a product in the scattering equations.  We quote a disk mass of $(2\pm0.5)\times10^{-3}$ $M_\sun$ for HOPS 136 with the understanding that it can be compared most directly to modeling done with similar opacity laws; we have chosen dust properties similar to widely used tabulations.  With a larger disk mass, the central lane in the image is too dark, as measured by the separation of the contours at 35\% of maximum brightness, and the 10 \micron\ minimum in the SED is too deep. The opposite problems arise for smaller disk masses.

The scale height at large radius controls the morphology of the bright nebulae just above and below the dark lane, which we track with the image contours at 10\% of maximum flux.  We quote a scale height at 100 AU of $12\pm2$ AU.  A larger scale height makes the 10\% contours too high and the dark lane too wide; the opposite problems arise for a smaller scale height.

\citet{wat07} caution that the radius corresponding to the edge of the scattered light from the disk is merely a lower limit to the disk radius.  From this approach, we find a disk radius of $450\pm25$ AU.  While this may be a lower limit, we note that it is already larger than most known protostellar disks.  We discuss this in further detail in Section~\ref{s.discussdisk}.

\subsection{Envelope Properties}

As discussed above, the SED and images respond to the overall envelope density, which can be quantified in various ways. We quote a reference density (Eqn.~\ref{e.rho1}) of $\rho_1= (4.5\pm0.5)\times10^{-15}$~g~cm$^{-3}$ for comparison with literature studies but note that densities as large as $\rho_1$ do not actually occur anywhere in the envelope.  To give a sense of a typical envelope density, $\rho_{1000}=\rho_1\times1000^{-1.5}$, the density at 1000 AU in the limit of no rotation (E.\ Furlan et al., in preparation), is $(1.4\pm0.2)\times10^{-19}$~g~cm$^{-3}$ for HOPS 136.  Larger envelope densities are ruled out because they yield a far-IR SED peak that is too red and near-IR images in which too little scattered light escapes the envelope.  Smaller envelope densities are ruled out because they yield a far-IR SED peak that is too weak and near-IR images in which the contours of the scattered light at 1\% of maximum emission are too concentrated toward the disk.

We assume the centrifugal radius $R_C$, where the dependence of the density on the stellocentric radius $r$ transitions from $r^{-0.5}$ when $r\ll R_C$ to $r^{-1.5}$ when $r\gg R_C$, is the same as the disk outer radius.

Given the \citet{ulr76} density law, our best-fit envelope falls to a density typical of the larger molecular cloud, $\sim10^3$ molecules cm$^{-3}$, at $\sim$ 10,000 AU.  We adopt this as the outer radius of the envelope.  Accounting for the reference envelope density $\rho_1$, the centrifugal radius $R_C$, and the cavity parameters described below, we calculate an envelope mass inside a 10,000 AU radius of $0.060\pm0.005$ $M_\sun$.

\subsection{Cavity Properties}

In Figure~\ref{f.nicmos}, the northern cavity is better defined than the southern one.  We measured the apparent (half) opening angle of the cavity at 2000 AU above the dark lane to be 21$^\circ$.  The cavity shape near its base is curved, suggestive of a parabola.  A parabolic cavity with an opening angle of 21$^\circ$ at 2000 AU corresponds to a 10$^\circ$ opening angle at the 10,000 AU envelope radius.  We thus quote an opening angle of $10\pm2^\circ$ for the cavity and an opening exponent of $2\pm0.5$.

An unusual aspect of the HOPS 136 SED (Fig.~\ref{f.sed}) is the redness of the short-wavelength peak (near 2.2~\micron) and the relatively small ratio of the long-wavelength peak to the short-wavelength peak, 5.5 in $\lambda S_\lambda$ space.  For comparison, in the SED generated by adopting the reference Class I parameters from \citet{whi03a} and using the dust opacities described above, the short-wavelength peak is at only 1.7 \micron\ and the ratio of the long-wavelength peak flux to the short-wavelength peak flux is 15.  These unusual characteristics of the HOPS 136 SED can be replicated by adding material to the envelope cavities. This is parameterized by setting the density at $R_*$ to $2.0\times10^{-13}$ g cm$^{-3}$, diminishing with radius to the $-1.5$ power.  In the model, no dust exists inside the sublimation radius ($14.8~R_*$), so the largest density encountered in the cavity is $(3.3\pm0.5)\times10^{-15}$ g cm$^{-3}$.  The total mass in each cavity is $(5.4\pm0.8)\times10^{-5}~M_\sun$, and the optical depth at 2.2 \micron\ through the cavity along the polar axis is $\tau_{2.2}=0.38\pm0.06$.  Along the cavity wall, the cavity is always much less dense than the adjacent disk or envelope, consistent with cavity dust of a reasonably low mass and density.  This cavity dust shifts the near-IR peak to the requisite longer wavelength and larger relative flux.

\section{DISCUSSION}

Working from the center of the system outward, we discuss the central star, the disk, and the envelope and cavity of HOPS 136.

\subsection{Estimating the Central Source Mass\label{s.star}}

Without a measurement of Keplerian rotation, the properties of the central source are uncertain in a protostellar system.  The weak flux in the optical and near-IR and the veiling of photospheric lines in these regions by accretion processes prevent spectral typing.  SED modeling alone is not useful, as one cannot distinguish between the intrinsic luminosity of the central star and the luminosity due to accretion, and, in our adopted model for the envelope density, the envelope infall rate and the mass of the star are degenerate.

Nonetheless, we can combine the results of SED fitting with envelope and stellar evolution models to estimate a plausible mass for the central source.  Averaged over a long period, the envelope infall rate for HOPS 136 is unlikely to be greater than $6\times10^{-7}$ $M_\sun~{\rm yr}^{-1}$, as this would deplete the modeled envelope mass of 0.06 $M_\sun$ in less than $10^5$ yr.  The current disk accretion rate is also unlikely to be significantly larger than this, as HOPS 136 shows no evidence of the well studied accretion outbursts in young stellar objects \citep[e.g.,][]{rei10}.

First, the estimated maximum envelope infall rate and modeled envelope density place a rough upper limit on the mass of the central object.  Equation~\ref{e.rho1} shows that, for a known envelope density, the envelope infall rate scales as the square root of the central mass.  An overly massive central star would cause the envelope to fall in implausibly quickly.  Substituting our finding of $\rho_1= 4.5\times10^{-15}$~g~cm$^{-3}$ and our estimate of $\dot{M}_{\rm env} \le 6\times10^{-7}$ $M_\sun~{\rm yr}^{-1}$ into Equation \ref{e.rho1}, we find $M_*\le 0.5~M_\sun$.  The total modeled luminosity of 4.7 $L_\sun$ is also indicative of a low-mass star.

Second, the estimated maximum disk accretion rate and total modeled luminosity place a rough lower limit on the mass of the central object.  The total modeled luminosity is the sum of stellar luminosity $L_*$ (due to contraction toward the main sequence) and accretion luminosity $L_{\rm acc}$.  (This neglects luminosity due to external heating, but this is expected to be insignificant in an isolated source such as HOPS 136.)  Essentially, our estimate of a low disk accretion rate suggests a relatively small accretion luminosity and a relatively large stellar luminosity, ruling out the least massive stars.

 The total luminosity is \begin{equation}
L=L_*+L_{\rm acc}=L_*+\epsilon~G M_* \dot{M}_{\rm disk} / R_*,
\end{equation} where $\epsilon$ is the fraction of the potential energy that is radiated away, and $G$ is the gravitational constant.  In accreting young stars, $\epsilon$ is commonly assumed to be 0.8 due to the truncation of the accretion disk at a few stellar radii by the stellar magnetic field \citep{gul98}, a value we adopt here.  Rearranging and substituting, \begin{eqnarray}
\dot{M}_{\rm disk} & = & 3.2\times10^{-8}\left(\frac{R_*/R_\sun}{M_*/M_\sun}\right)\!\left(\frac{L-L_*}{L_\odot}\right)\!\left(\frac{1}{\epsilon}\right) M_\odot\;{\rm yr}^{-1} \nonumber \\ & \le & 6\times10^{-7}~M_\sun~{\rm yr}^{-1}.
\label{e.mdot}\end{eqnarray}

Further rearranging terms, replacing $L$ and $\epsilon$ with our favored values, and using solar units, we find  \begin{equation}\frac{M_*}{R_*}+0.067~L_* \ge 0.31.\end{equation}  This condition can be compared to models of pre-main-sequence stars, since the small remaining envelope mass suggests the main accretion phase is over. Although the model age to assign to HOPS 136 is a source of ambiguity, inspection of the \citet{sie00} models for small ages reveals that the left-hand side of this condition generally increases with $M_*$, exceeding the right-hand side at about 0.4 $M_\sun$ for a model age of $10^4$ yr after the end of the main accretion phase.  We thus adopt a lower stellar mass limit of $\sim0.4$ $M_\sun$. With the upper limit of 0.5~$M_\sun$ from Equation~\ref{e.rho1}, we estimate that HOPS~136 contains a star of $\sim0.4-0.5$ $M_\sun$. This is typical of low-mass protostars but substantially in excess of the 0.06~$M_\sun$ remaining in the envelope.

\vskip 0.5 in

\subsection{The Disk of HOPS 136 in Context\label{s.discussdisk}}

The mass and radius of the HOPS 136 disk can be compared to those of large samples of disks from Class 0 to Class II.  Table~\ref{t.masses} compares findings from several studies to our results for HOPS 136.  \citet{jor09} observed 20 Class 0 and I protostars between 850 \micron\ and 1.3 mm.  \citet{eis12} imaged ten Class I objects in Taurus at 1.3 mm, detecting eight single sources and one binary, and modeled these images in concert with 0.9 \micron\ images and broadband SEDs from the literature.  (Table~\ref{t.masses} excludes the binary.) \citet{gra10} took an approach similar to ours, modeling the 0.55 to 1300 \micron\ SEDs and near-to-mid-IR images of eight Class I objects in Taurus.  Focusing on later stages, \citet{and05} presented masses for 16 Class I, 9 flat-SED, and 64 Class II disks in Taurus based on 850 \micron\ fluxes.  Finally, the online Catalog of Circumstellar Disks\footnote{\url{http://www.circumstellardisks.org/}} \citep{wat07} lists diameters for 16 Class I objects (category YSO) and 54 Class~II objects (category TT).

\begin{deluxetable}{lcccc}
\tablecaption{Disk Properties\label{t.masses}}
\tablewidth{\hsize}
\tablehead{\colhead{Reference} & \colhead{Class} & \colhead{$M_{\rm disk}$ ($M_\sun$)} & \colhead{$M_{\rm env}/M_{\rm disk}$} & \colhead{$R_{\rm disk}$ (AU)}}
\startdata
J09 & 0 & 0.089 & 19 & \nodata \\
J09 & I & 0.011 & 4 & \nodata \\
E12 & I & 0.008 & 12 & 310 \\
CD & I & \nodata & \nodata & 190 \\
G10 & I & 0.010 & 10 & 120 \\
A05 & I & 0.030 & \nodata & \nodata \\
A05 & FS & 0.004 & \nodata & \nodata \\
A05 & II & 0.003 & \nodata & \nodata \\
CD & II & \nodata & \nodata & 110 \\
HOPS 136 & I & 0.002 & 30 & 450
\enddata
\tablecomments{Results from compilations are medians. FS refers to flat-SED sources.}
\tablerefs{(J09) \citealt{jor09}; (E12) \citealt{eis12}; (G10) \citealt{gra10}; (A05) \citealt{and05}; (CD) Catalog of Circumstellar Disks}
\end{deluxetable}

By several observational diagnostics (Section 3) and our modeling (Section 4), HOPS 136 is a Class I object, corresponding to the stage of protostellar evolution in which an envelope is still falling onto a circumstellar disk.  However, its disk mass of 0.002 $M_\sun$ is more typical of Class II objects, where the envelope has essentially dissipated.  Its mass is below the median mass of the \citet{and05} Class II disks, in the 40th percentile.  Its envelope mass is small for Class I, but its ratio of envelope mass to disk mass, 30, is large compared to the median ratios of the \citet{jor09}, \citet{eis12}, and \citet{gra10} embedded samples.

The radius of the disk, 450 AU, is large for disks at any stage.  It exceeds the median radius for the \citet{eis12} sample and is equal to those of its two largest disks.   It is also larger than those of all objects in the \citet{gra10} sample except the ``Butterfly Star'' IRAS 04302+2247, a similarly edge-on protostar with a disk radius of 500 AU.  In the Catalog of Circumstellar Disks, HOPS 136 is in the 80th percentile of the Class I radius distribution and the 90th percentile of the Class II distribution.

To summarize, the mass of the HOPS 136 disk is typical of Class II objects, but consistent with its classification as a Class I protostar, there is a substantial reservoir of mass remaining in its envelope.  Its radius is large (but not unprecedented) for either a Class I or a Class II system, suggesting that the infalling envelope may have imparted a large amount of angular momentum to the disk.  It appears to be an example of a system in transition from the protostellar stage to the T Tauri stage.

\subsection{HOPS 136: Approaching the End of Envelope Infall}

We showed in Section~\ref{s.source} how, from its near- and mid-IR photometry and spectrum alone, one might draw alternative conclusions about the evolutionary state of HOPS 136.  The slope of the SED across the {\em Spitzer}/IRAC range and the $K$ to 24 \micron\ slope both indicate that HOPS 136 is a flat-SED source.  The {\em Spitzer}/IRS classification scheme of \citet{mcc10} identifies HOPS 136 as a transition disk.  The SEDs of edge-on protostars and face-on transition disks are similar in that both have mid-IR deficits.  In transition disks, optically thick material with the requisite temperatures for mid-IR emission is not present, while in edge-on protostars, this material is hidden from view by the cold outer regions of the disk.

With far-IR photometry and HST imaging, HOPS 136 is revealed to retain a protostellar envelope.  It has $\alpha_{2.2-70}=0.5$, making it a Class I source, and the imaging shows light scattered by the inner envelope.   Calculating $\alpha_{2.2-70}$ for the best-fit model to HOPS 136 (as opposed to the observed photometry) has the advantage that we can estimate the dependence of this quantity on inclination angle.  The near-to-far-IR slope of the model SED is consistent with Class I sources ($\alpha_{2.2-70}\ge0.3$) when the inclination is closer to edge-on than $70^\circ$ and consistent with flat-SED sources ($-0.3\le \alpha_{2.2-70}<0.3$) when the inclination is closer to pole-on than $70^\circ$.  This is evidence that flat-SED and some Class I sources may be objects in a similar evolutionary state viewed from different angles.

The inclination dependence of $L_{\rm smm}/L_{\rm bol}$ is also a clue to the evolutionary state of HOPS 136.  The observed ratio is 0.8\%, which indicates a dense Class 0 envelope under the original \citet{and93} definition.  However, this ratio is enhanced by the edge-on inclination.  As the inclination of the best-fit model varies from edge-on to pole-on, $L_{\rm smm}$ increases by only 7\%, but $L_{\rm bol}$ increases by nearly 700\%.  This is because the disk attenuation that so strongly affects the edge-on SED predominantly reduces the fluxes at wavelengths less than 160~\micron.  Thus, $L_{\rm smm}/L_{\rm bol}$ is greater than the \citet{and93} Class~I upper limit of 0.5\% only for inclination angles greater than about $80^\circ$, again pointing to a protostellar envelope in the later stages of evolution.

\subsection{The Origin of the Cavity Material}

Simultaneously fitting the SED and near-IR images of HOPS 136 requires a small amount of dust and gas in the cavity. While \citet{whi03a} set a constant cavity density of $1.67\times10^{-19}$ g cm$^{-3}$ in their Class~I model, which would correspond to $0.014~M_\sun$ of material in each of the adopted HOPS 136 cavities, this low density has an undetectable effect on our model SED and images.  Our prescription for the cavity density adds negligible mass ($5.4\times10^{-5}~M_\sun$ per cavity); rather, the improved SED fit is due to the concentration of cavity material near the disk plane.  The density averaged over the inner 1 AU of each cavity is $2.9\times10^{-16}$ g cm$^{-3}$, three orders of magnitude larger than the constant density of \citet{whi03a} and comparable to the constant cavity density used by \citet{gra10} to model the Butterfly Star.  (This is the only one of the eight embedded objects modeled by \citealt{gra10} that requires cavity dust of sufficient density to affect the SED, which weakly constrains the frequency of this phenomenon.)  Our $1/r^{1.5}$ density law yields the desired behavior in the SED without requiring a substantial increase in the cavity mass.  

\citet{zha13} showed how, in massive stars, a disk wind in the style of \citet{bla82} or an X-wind in the style of \citet{shu94} could carry disk dust into an outflow cavity.  The same process may be at work in HOPS 136.

\section{CONCLUSIONS}

In {\em Spitzer} and {\em Hubble} images of the L~1641 region in the Orion A molecular cloud, we discovered an isolated edge-on protostar and designated it HOPS 136 in the target catalog for the {\em Herschel} Orion Protostar Survey.  In NICMOS 1.60 and 2.05 \micron\ images, the object is almost exactly edge-on and shows all of the hallmark features of an embedded protostar:\ a dark dust lane and bright scattered light from the disk, diffuse scattered light from an infalling envelope interrupted by a disk shadow, and bipolar cavities presumably evacuated by past outflow.

We used the \citet{whi03b} radiative transfer code to model the {\em HST} images and the $1-1200$~\micron\ SED of the source, which combines photometry and spectra from 2MASS, {\em WISE}, {\em Spitzer}, {\em Herschel}, APEX, and IRAM.  From the modeling, we conclude that the protostar is of moderate luminosity ($L=4.7~L_\sun$) and envelope mass (0.06~$M_\sun$ inside 10,000 AU).  Its disk has mass 0.002 $M_\sun$ and radius 450 AU, and there is $5.4\times10^{-5}~M_\sun$ of material in each envelope cavity concentrated near the disk plane.  With pre-main sequence models, we estimate a central source mass between 0.4 and 0.5 $M_\sun$.

By the observational diagnostics $T_{\rm bol}$ and $\alpha_{2.2-70}$, HOPS 136 is a Class I source.  With modeling, we find the envelope mass to be much less than the stellar mass but not zero, confirming that it is approaching the end of envelope infall.  Its disk mass, however, is less than the median for T Tauri stars, and from most inclination angles, HOPS 136 is expected to resemble a flat-SED source.  HOPS 136 appears to be an example of a system in transition from the protostellar stage to the T Tauri stage, when disk conditions are being established that later determine the architectures of planetary systems.

\acknowledgments

Support for this work was provided by NASA through awards issued by JPL/Caltech. This paper includes data from the {\em Spitzer Space Telescope}, which is operated by JPL/Caltech under a contract with NASA.  We also include observations made under program 11548 of the NASA/ESA {\em Hubble Space Telescope}, obtained at the Space Telescope Science Institute, which is operated by the Association of Universities for Research in Astronomy, Inc., under NASA contract NAS 5-26555.

We include data from the Atacama Pathfinder Experiment, a collaboration between the Max-Planck-Institut f\"{u}r Radioastronomie, the European Southern Observatory, and the Onsala Space Observatory.  These data were collected at the European Southern Observatory, Chile, under proposal 088.C-0994. This paper makes use of data products from the Two Micron All Sky Survey, which is a joint project of the University of Massachusetts and the Infrared Processing and Analysis Center, Caltech, funded by NASA and the National Science Foundation (NSF), and the {\em Wide-field Infrared Survey Explorer}, which is a joint project of the University of California, Los Angeles, and JPL/Caltech, funded by NASA.

JJT acknowledges support provided by NASA through Hubble Fellowship grant \#HST-HF-51300.01-A awarded by the Space Telescope Science Institute, which is operated by the Association of Universities for Research in Astronomy,  Inc., for NASA, under contract NAS 5-26555. The National Radio Astronomy Observatory is a facility of the National Science Foundation operated under cooperative agreement by Associated Universities, Inc.

The work of AMS was supported by the Deutsche Forschungsgemeinschaft priority program 1573 (``Physics of the Interstellar Medium''). MK acknowledges support from the NSF Research Experiences for Undergraduates program of the Department of Physics and Astronomy at the University of Toledo (grant PHY-1004649).  MO acknowledges support from MINECO (Spain) AYA2011-30228-C03 grant (co-funded with FEDER funds). We are grateful to Barbara Whitney, Tom Robitaille, and their collaborators for making their radiative transfer code and fitting tools available to the community.

\end{document}